\documentclass[preprint,showpacs,preprintnumbers,amsmath,amssymb]{revtex4}

\usepackage{graphicx}
\usepackage{dcolumn}
\usepackage{bm}

\begin{document}

\title{Dispersive properties of quasi-phase-matched optical parametric amplifiers}

\author{S. Longhi, M. Marano, and P. Laporta}

\affiliation{%
\\
INFM, Dipartimento di Fisica, Politecnico di Milano, Piazza L. da Vinci, 32, I-20133
Milano (Italy) }%


\begin{abstract}
The dispersive properties of non-degenerate optical parametric amplification in
quasi-phase-matched (QPM) nonlinear quadratic crystals with an arbitrary grating
profile are theoretically investigated in the  no-pump-depletion limit. The spectral
group delay curve of the amplifier is shown to be {\it univocally} determined by its
spectral power gain curve through a Hilbert transform. Such a constraint has important
implications on the propagation of spectrally-narrow optical pulses through the
amplifier. In particular, it is shown that anomalous transit times, corresponding to
superluminal or even negative group velocities, are possible near local minima of the
spectral gain curve. A possible experimental observation of such effects using a QPM
Lithium-Niobate crystal is suggested.

\end{abstract}

\maketitle

\section{\label{intro} Introduction }
The theoretical and experimental study of second-order nonlinear processes has received
in the past years a renewed interest after the introduction of the quasi-phase-matching
(QPM) technique \cite{Armstrong62,Fejer92}, which has lead to a major advance in
applications such as frequency conversion, parametric oscillation and amplification,
nonlinear optical frequency mixing and pulse shaping (see, for instance,
\cite{Pierce97,Byer97,Fejer98} and references therein). The QPM technique uses a
periodic modulation of the nonlinear $\chi^{(2)}$ coefficient (QPM grating) to
compensate for refractive-index dispersion, permitting phase matching operation at any
wavelength at room temperature, which would not be possible with birefringent phase
matching. In addition, QPM enables the use of materials with strong nonlinearities
which are not phase matchable by angle or temperature tuning, with a polarization state
of interacting fields corresponding to the largest diagonal element of the $\chi^{(2)}$
tensor. Though the physics of the QPM technique is known since a long time
\cite{Armstrong62}, only with the recent technological advances in the electric-field
poling of ferroelectric materials, such as LiNbO$_3$, LiTaO$_3$ and KTP, the
experimental potentialities of QPM-based devices have become practicable and have
reached nowadays a satisfactory degree of maturity \cite{Fejer98}. The possibility of
engineering a QPM grating by breaking its periodicity introduces additional degrees of
freedom for light control which open the way for a novel class of devices
\cite{Arbore97a,Arbore97b,Mizuuchi98,Imeshev98,Chou99,Bang99,Imeshev00,Imeshev01,Liu01,Zhang01}.
In particular, pulse compression and shaping by second harmonic generation (SHG) or by
difference frequency generation (DFG) have been proposed and demonstrated using
nonuniform QPM gratings \cite{Arbore97a,Arbore97b,Imeshev98,Imeshev00,Imeshev01}. In
such cases, it was shown that in the low-conversion regime the spectral transfer
function governing the frequency conversion process is related to the longitudinal QPM
grating profile by a simple Fourier transform (for a detailed analysis and for a
discussion of the distinctive aspects of SHG and DFG cases, see
\cite{Imeshev00,Imeshev01}). The use of a QPM nonlinear crystal for parametric
amplification of a signal field at frequency $\omega_1$ by a strong pump field at
frequency $\omega_3$ has been demonstrated as well, both in waveguide and bulk
geometries \cite{Bortz95,Galvanauskas98}, with the achievements of gains as large as 40
dB using nanosecond pump pulses \cite{Galvanauskas98}. As compared to most common
parametric amplifiers based on birefringent phase matching, the use of a QPM-based
parametric amplifier may offer the possibility of molding the spectral gain response of
the amplifier by a suitable design of the QPM grating (see, e.g., \cite{Zhang01}).
Though the process of parametric amplification is accompanied by the generation of the
idler field at frequency $\omega_2=\omega_3-\omega_1$ and is thus equivalent to the DFG
scheme recently studied in literature \cite{Imeshev01,Liu01}, the simple Fourier
analysis developed for DFG \cite{Imeshev01}, which neglects amplification of the signal
field, is inadequate to study the parametric amplification process in QPM gratings, and
a more complex analysis is needed, based e.g. on matrix transfer techniques
\cite{Zhang01}. A general theoretical analysis of the properties of a QPM parametric
amplifier, and especially of the interplay between its gain and dispersive properties,
has not been pursued yet. Such a problem is closely related to the "inverse problem" of
QPM grating synthesis, i.e. of the determination of a QPM grating profile that realizes
a desired spectral gain response
of the amplifier.\\
In this paper we provide a general analysis of the process of parametric amplification
in an aperiodic QPM nonlinear crystal. We will show in a very general way that the
spectral power gain curve of the amplifier univocally determines its dispersive
properties, whereas the QPM grating profile that realizes such a spectral gain curve is
not uniquely determined. The interplay between the spectral power gain and dispersive
curves of the amplifier may lead to abnormal dispersive properties of the amplifier,
such as the occurrence of superluminal or even negative group velocities similar to
those observed in atomic amplifiers with a gain doublet
\cite{Steinberg94,Wang00,Dogariu01}. The paper is organized as follows. In Sec.II the
basic equations describing parametric pulse amplification in a QPM nonlinear crystal
are reviewed. In the undepleted pump approximation, general properties of spectral gain
and dispersion curves are derived in Sec.III, and the inverse problem of QPM synthesis
is addressed. Section IV deals with the propagation of spectrally narrow pulses and
analyzes the occurrence of anomalous regions corresponding to superluminal or even
negative group velocities. Finally, in Sec.V the main conclusions are outlined.

\section{\label{model} Parametric amplification in QPM quadratic media: basic equations}
The starting point of the analysis is provided by the propagation scalar wave equation
for a linearly-polarized electric field ${\cal E}(z,t)$ in a nonlinear $\chi^{(2)}$
medium with a quasiperiodic QPM grating profile. Under the usual plane-wave
approximation and taking into account material dispersion, from Maxwell's equations one
can write (see, for instance, \cite{Newell92})
\begin{equation}
\frac{\partial^2 {\cal E}}{\partial z^2}+ \int_{-\infty}^{\infty} d \omega k^2(\omega)
\tilde {\cal E}(z,\omega) \exp(-i \omega t) =\mu_0 \frac{\partial^2 {\cal
P}^{NL}}{\partial t^2} \; ,
\end{equation}
where $\tilde {\cal E}(z,\omega)=(2 \pi)^{-1} \int_{-\infty}^{\infty} d\omega {\cal
E}(z,t) \exp(i \omega t)$ is the Fourier transform of ${\cal E}(z,t)$,
$k(\omega)=(\omega / c_0) \sqrt {1 + \tilde \chi(\omega)}=(\omega/c_0)n(\omega)$ is the
dispersion relation defined by the complex linear susceptibility $\tilde {\chi}
(\omega)$ [or by the complex refractive index $n(\omega)= \sqrt{1+\tilde
{\chi}(\omega)}$], $c_0$ is the speed of light in vacuum, $\mu_0$ is the vacuum
magnetic permeability, and ${\cal P}^{NL}$ is the nonlinear driving polarization term.
For a quadratic medium and neglecting dispersion effects of second-order polarization,
one can take ${\cal P}^{NL}(z,t)=\epsilon_0 \chi^{(2)}(z) {\cal E}^2(z,t)$, where
$\chi^{(2)}$ is the spatially-modulated nonlinear susceptibility that accounts for the
QPM grating. To study parametric amplification or DFG, we assume that ${\cal E}(z,t)$
is described by the superposition of three wavetrains with carrier frequencies
$\omega_1$, $\omega_2$ and $\omega_3=\omega_1+\omega_2$, corresponding to the signal,
idler and pump waves, respectively. We thus set
\begin{equation}
{\cal E}(z,t)= \frac{1}{2}\left[ {\cal E}_1(z,t) \exp(-i \omega_1 t)+ {\cal E}_2 (z,t)
\exp(-i\omega_2 t)+{\cal E}_3(z,t) \exp(-i\omega_3 t) +c.c. \right] \; ,
\end{equation}
where the envelopes ${\cal E}_{1,2,3}(z,t)$ of wavetrains are assumed to vary slowly
with respect to time $t$ as compared to the exponential terms. In case of parametric
amplification (or DFG), a strong pump wave and a weak signal wave are incident upon the
nonlinear medium at the entrance plane $z=0$, whereas the idler wave is generated by
parametric interaction in the nonlinear medium and appears at the output plane $z=L$
(see Fig.1). Substitution of Eq.(2) into Eq.(1) and setting equal the terms oscillating
at the same frequency, yields the following set of coupled-wave equations:
\begin{subequations}
\begin{eqnarray}
\frac{\partial^2 {\cal E}_1}{\partial z^2}+k^2(\omega_1+i \partial_t) {\cal
E}_1=-\chi^{(2)} \left( \frac{\omega_1}{c_0} \right)^2 {\cal E}_3 {\cal E}_{2}^{*} \\
\frac{\partial^2 {\cal E}_2}{\partial z^2}+k^2(\omega_2+i \partial_t) {\cal
E}_2=-\chi^{(2)} \left( \frac{\omega_2}{c_0} \right)^2 {\cal E}_3 {\cal E}_{1}^{*} \\
\frac{\partial^2 {\cal E}_3}{\partial z^2}+k^2(\omega_3+i \partial_t) {\cal
E}_3=-\chi^{(2)} \left( \frac{\omega_3}{c_0} \right)^2 {\cal E}_1 {\cal E}_2
\end{eqnarray}
\end{subequations}
In deriving Eqs.(3), we have neglected the nonresonant terms in the nonlinear
polarization driving term ${\cal P}^{NL}$, and we used the following property (see, for
instance, \cite{Newell92}):
\begin{equation}
\int_{-\infty}^{\infty} d \omega k^2(\omega) \tilde {\cal E}(\omega) \exp(-i \omega t)
= \left[ k^2(\omega_0+i \partial_t ) {\cal A}(t)\right] \exp(-i \omega_0 t)
\end{equation}
which is valid for any signal of the form ${\cal E}(t)={\cal A}(t) \exp(-i \omega_0
t)$, where the operator $k^2(\omega_0+i \partial_t)$ is defined through its power
series expansion:
\begin{equation}
k^2(\omega_0+i \partial_t) \equiv k^2(\omega_0) +  2 i k(\omega_0) \left(
\frac{\partial k}{\partial \omega} \right)_{\omega_0} \frac{\partial}{\partial t} -
\left[ \left( \frac{\partial k}{\partial \omega} \right)^{2}_{\omega_0}+k(\omega_0)
\left( \frac{\partial^2 k}{\partial \omega^2} \right)_{\omega_0}  \right]
\frac{\partial^2}{\partial t^2}+ ...
\end{equation}
Notice that, in case of a wavetrain at carrier frequency $\omega_0$ and slowly-varying
envelope ${\cal A}(t)$, the power series on the right hand side in Eq.(5) may be
truncated at some order. At leading order in the expansion one retrieves the wavenumber
at the carrier frequency $\omega_0$, the second one determines the group-velocity
$v_g=1/(\partial k /
\partial \omega)$ of the wavetrain, whereas higher order terms account for group
velocity dispersion. If in Eqs.(3) the nonlinear terms are assumed to be weak and we
truncate expansion (5) keeping the first term only, one sees that at leading order the
amplitudes ${\cal E}_{l}$ are oscillating in space like $\exp( \pm i k_l z)$, where
$k_l \equiv k(\omega_l)$ ($l=1,2,3)$. If we consider only forward propagating waves,
 it is convenient to remove the fast
oscillation over the wavelength spatial scale by setting ${\cal E}_l(z,t)=A_l(z,t)
\exp(i k_lz)$. For the QPM grating, we assume that $\chi^{(2)}$ is a quasi-periodic
function of $z$ with period $\Lambda$, i.e. :
\begin{equation}
\chi^{(2)}(z)= \sum_{n=-\infty}^{\infty} \chi^{(2)}_{n}(z) \exp(-2 i n \pi z / \Lambda)
\; ,
\end{equation}
where the Fourier coefficients ${\chi^{(2)}_{n}(z)}$ are slowly varying functions of
$z$ over one period $\Lambda$. In practice, the slow dependence of coefficients on $z$
can be achieved by a +/- reversal of domains in the ferroelectric crystal with a local
period and local duty cycle that are slowly varying along the $z$ axis (see, for
instance, \cite{Imeshev00}). We assume that the nominal QPM period satisfies the phase
matching condition:
\begin{equation}
\Lambda=M \frac{2 \pi}{\Delta k} \; ,
\end{equation}
where $\Delta k \equiv k_3-k_2-k_1$ is the wavevector mismatch of interacting waves and
$M$ is a positive integer (QPM of order $M$). The evolution equations for the envelopes
$A_l(z,t)$ ($l=1,2,3$) can be derived, in the limit of weak nonlinearity and
quasi-monochromatic wavetrains, by a multiple-scales asymptotic expansion (see, for
instance, \cite{Bang97}); for the sake of completeness, a brief account of the
derivation of the envelope equations is given in Appendix A. These reads:
\begin{subequations}
\begin{eqnarray}
2ik_1 \frac{\partial A_1}{\partial z}=\left[ k_{1}^{2}-k^2(\omega_1+i \partial_t)
\right] A_1-\frac{2k_{1}^{2}}{n_{1}^{2}}d_{eff}A_{2}^{*}A_3 \\
2ik_2 \frac{\partial A_2}{\partial z}=\left[ k_{2}^{2}-k^2(\omega_2+i \partial_t)
\right] A_2-\frac{2k_{2}^{2}}{n_{2}^{2}}d_{eff}A_{1}^{*}A_3 \\
2ik_3 \frac{\partial A_3}{\partial z}=\left[ k_{3}^{2}-k^2(\omega_3+i \partial_t)
\right] A_3-\frac{2k_{3}^{2}}{n_{3}^{2}}d_{eff}^{*} A_1A_2
\end{eqnarray}
\end{subequations}
where:
\begin{equation}
d_{eff}(z) \equiv \frac{1}{2} \overline {\chi^{(2)}(z) \exp(i \Delta k z) }=
\frac{1}{2} \chi^{(2)}_{M}
\end{equation}
and the overline denotes a spatial average over a few modulation periods of the QPM
grating. Equations (8) are the basic equations describing QPM parametric processes in
the time domain. The linear operators on the right hand side of Eqs.(8), describing the
linear dispersive and absorptive properties of the medium, may be expanded in power
series [see Eq.(5)], and the number of terms that need to be kept depends on the the
spectral extension of the wavetrains. In the following, we will consider spectral
regions of transparency for the medium, so that we will neglect the imaginary part of
the dispersion relation $k(\omega)$. In addition, for spectrally-narrow envelopes, only
the leading-order term in the expansion may be kept, and Eqs.(8) reduce to the
following ones:
\begin{subequations}
\begin{eqnarray}
\frac{\partial A_1}{\partial z}+\frac{1}{v_{g1}} \frac{\partial A_1}{\partial t}=
i \frac{k_{1}}{n_{1}^{2}}d_{eff}A_{2}^{*}A_3 \\
\frac{\partial A_2}{\partial z}+\frac{1}{v_{g2}} \frac{\partial A_2}{\partial t}=
i \frac{k_{2}}{n_{2}^{2}}d_{eff}A_{1}^{*}A_3 \\
\frac{\partial A_3}{\partial z}+\frac{1}{v_{g3}} \frac{\partial A_3}{\partial t}= i
\frac{k_{3}}{n_{3}^{2}}d_{eff}^{*}A_{1}A_2
\end{eqnarray}
\end{subequations}
where $v_{gl}=1/(\partial k / \partial \omega )_{\omega_l}$ ($l=1,2,3$) are the group
velocities of the three wavetrains. For spectrally broad wavetrains, higher-order terms
in the expansion (5) should be considered, which are responsible for group velocity
dispersion and higher-order dispersion effects.

\section{\label{dispersion} Spectral gain curve and group delay analysis}
In this section we provide a general analytical framework to determine the complex
spectral gain curve of a QPM parametric amplifier in the limit of undepleted pump and
derive a general relation among dispersive and gain properties of the amplifier. We
assume that the amplifier is pumped by a continuous-wave (CW) or quasi CW strong pump
wave at frequency $\omega_3$, and a weak signal field at carrier frequency $\omega_1$,
described by the envelope $A_1(0,t)= \int_{-\infty}^{\infty} \tilde f(\Omega) \exp(-i
\Omega t)$, is incident upon the nonlinear crystal at the plane $z=0$. In the limit of
no pump depletion, which is valid for a low or moderate pump conversion, we may
disregard Eq.(8c) and assume in Eqs.(8a) and (8b) a constant value of $A_3$, given by
$A_3=[2I_3/(\epsilon_0 c_0 n_3)]^{1/2}$, where $I_3$ is the intensity of the incident
CW pump and $n_3=n(\omega_3)$. In this case, Eqs.(8a) and (8b) become linear equations,
and the signal field envelope $A_1(L,t)$ at the exit of the medium can be written using
standard Fourier analysis in the form:
\begin{equation}
A_1(L,t)= \int_{-\infty}^{\infty} d \Omega g(\Omega) \tilde f (\Omega) \exp(-i \Omega
t)
\end{equation}
where $\tilde f (\Omega)$ is the Fourier transform of the incident signal wavetrain and
$g(\Omega)$ is the spectral gain curve of the amplifier. In order to determine the
spectral gain curve $g(\Omega)$, let us search for a solution of Eqs.(8a) and (8b) in
the form:
\begin{subequations}
\begin{eqnarray}
A_1(z,t)& = & u(z) \exp[-i \Omega t +i \beta(\Omega)z] \\
A_2(z,t)& = & v^*(z) \frac{n_1}{n_2} \sqrt {\frac{k_2}{k_1}} \exp[i \Omega t -i \beta
(\Omega) z] \; ,
\end{eqnarray}
\end{subequations}
where $\Omega$ is the frequency offset from the reference carrier frequency $\omega_1$
of the signal wave, $n_1=n(\omega_1)$, $n_2=n(\omega_2)$, and $\beta=\beta(\Omega)$ is
defined by the relation:
\begin{equation}
\beta(\Omega) \equiv \frac{1}{4k_1} \left[ k^2(\omega_1+\Omega)-k_{1}^{2}
\right]-\frac{1}{4k_2} \left[k^2(\omega_2-\Omega) -k_{2}^{2} \right].
\end{equation}
The evolution equations of the complex envelopes $u(z)$ and $v(z)$ are obtained after
inserting the Ansatz (12) into Eqs.(8). One then obtains for $u(z)$ and $v(z)$ the
following coupled-mode equations:
\begin{subequations}
\begin{eqnarray}
\frac{du}{dz} & = & i \delta u +i q(z) v \\
\frac{dv}{dz} & = & -i \delta v -i q^*(z) u
\end{eqnarray}
\end{subequations}
where $\delta=\delta(\Omega)$ and $q(z)$ are give by:
\begin{equation}
\delta(\Omega)=\frac{1}{4k_1} \left[ k^2(\omega_1+\Omega)-k_{1}^{2} \right]
+\frac{1}{4k_2} \left[ k^2(\omega_2-\Omega)-k_{2}^{2} \right],
\end{equation}
\begin{equation}
q(z)= 2 \pi d_{eff}(z) \sqrt{\frac{2 I_3}{\epsilon_0 c_0 n_1 n_2 n_3 \lambda_1
\lambda_2 }}.
\end{equation}
In Eq.(16), $\lambda_1$ and $\lambda_2$ are the wavelengths in vacuum of signal and
idler fields. Equations (14) have the form of the Zakharov-Shabat system encountered in
problems of inverse scattering \cite{Ablowitz78} and grating analysis \cite{Poladian96,
Sipe94,Erdogan97,Othonos99}, the QPM grating profile $d_{eff}(z)$ playing the role of
the scattering potential [see Eq.(16)]. The general solution to Eqs.(14) can be written
as $ \left( u(L,\delta),v(L,\delta) \right)^T={\cal M} \left( u(0,\delta),v(0,\delta)
\right)^T$, where the elements of the $2 \times 2$ transfer matrix ${\cal M}={\cal
M}(\delta)$ satisfy the conditions ${\cal M}_{22}={\cal M}_{11}^{*}$, $ {\cal
M}_{21}={\cal M}_{12}^{*}$, and ${\rm det} {\cal M}={\cal M}_{11}{\cal M}_{22}-{\cal
M}_{12} {\cal M}_{21}=1$. From Eq.(12a), one realizes that the spectral gain curve
$g(\Omega)$ of the amplifier is given by \cite{note0}:
\begin{equation}
g(\Omega)= {\cal M}_{11}(\Omega) \exp[i \beta(\Omega) L ]
\end{equation}
where the dependence of ${\cal M}_{11}$ on $\Omega$ is through the function
$\delta=\delta(\Omega)$ [see Eq.(15)]. The explicit calculation of ${\cal M}(\delta)$
for a generic scattering potential $q(z)$, i.e. QPM grating profile $d_{eff}(z)$, can
be done numerically by a standard cascading technique in which the grating is
decomposed as the cascade of successive uniform sections and ${\cal M}$ is derived as
the product of the elementary matrices of each uniform grating section (see, for
instance, \cite{Erdogan97}). The application of the cascading technique for the study
of DFG and parametric amplification in certain nonuniform QPM gratings has been
recently addressed in \cite{Liu01,Zhang01}. The determination of the spectral gain
curve of a QPM parametric amplifier for a given QPM profile is thus easy: first, one
has to compute the matrix element ${\cal M}_{11}$ associated with the scattering
problem [Eqs.(14)] by, e.g., the transfer matrix technique, for different values of
frequency $\Omega$; the complex spectral gain curve is then retrieved by Eq.(17) with
the use of Eq.(13). Here we are mainly concerned on investigating the interplay between
amplitude and phase relationship of the complex spectral gain curve, which can be
derived in a very general way using the results of the inverse scattering theory. To
this aim, it is worth observing that, from inverse scattering theory, the scattering
coefficient $a(\delta) \equiv {\cal M}_{11}(\delta) \exp(-i \delta L) $ is an analytic
function in the lower part of the complex $\delta$ plane, i.e. for ${\rm Im}(\delta)
\leq 0$, $ |a(\delta)| \geq 1$ on the real axis, and $ a(\delta) \rightarrow 1$ as
$\delta \rightarrow \infty$ \cite{Ablowitz78}. In addition $a(\delta)$ has no zeros for
${\rm Im}(\delta) \leq 0$, so that $1/a(\delta)$ belongs to the class of causal and
minimal phase shift functions (see, for instance, \cite{Toll56}). This means that the
knowledge of the modulus of $a(\delta)$ univocally determines its phase. In fact, since
the function $F(\delta)=(\partial a /
\partial \delta) /a = \partial ({\rm ln} a) /
\partial \delta $ is analytic in the lower part of the complex $\delta$
plane and goes to zero for $\delta \rightarrow \infty$, the real
and imaginary parts of $F(\delta)$ can be related by a Hilbert
transform, that is:
\begin{equation}
F(\delta)=\frac{1}{\pi i} \int_{-\infty}^{\infty} d \delta'
\frac{F(\delta')}{\delta-\delta'}.
\end{equation}
After setting $a(\delta)=|a(\delta)| \exp[i \phi_a(\delta)]$ and making equal the
imaginary parts of both sides in Eq.(18), one obtains:
\begin{equation}
\frac{\partial \phi_a}{\partial \delta}= -\frac{1}{\pi } \int_{-\infty}^{\infty} d
\delta'  \frac{\partial {\rm ln}|a(\delta')|}{\partial \delta'}
\frac{1}{\delta-\delta'}
\end{equation}
In order to further proceeds in the analysis, we must specify the
dependence of $\beta$ and $\delta$ on $\Omega$. From Eqs.(13) and
(15), the expansion of $k=k(\omega)$ in in power series of
$\Omega$ yields:
\begin{equation}
\beta(\Omega)=\frac{1}{2} \Omega \left( \frac{1}{v_{g1}}+\frac{1}{v_{g2}}
\right)+\frac{1}{4} \Omega^2 \left(
k_{1}^{''}-k_{2}^{''}+\frac{k_{1}^{'2}}{k_1}-\frac{k_{2}^{'2}}{k_2} \right) +...
\end{equation}

\begin{equation}
\delta(\Omega)=\frac{1}{2} \Omega \left( \frac{1}{v_{g1}}-\frac{1}{v_{g2}}
\right)+\frac{1}{4} \Omega^2
\left(k_{1}^{''}+k_{2}^{''}+\frac{k_{1}^{'2}}{k_1}+\frac{k_{2}^{'2}}{k_2} \right) +...
\end{equation}
where $k^{'}_{1,2}$ and $k^{''}_{1,2}$ are the first and second derivatives of
$k(\omega)$ evaluated at the frequencies $\omega_{1,2}$. For the sake of simplicity,
let us assume that $v_{g1} \neq v_{g_2}$, which is usually the case for a
non-degenerate interaction, and that the bandwidth of the amplifier is narrow enough to
neglect group-velocity dispersion and higher-order dispersive effects in Eqs.(20) and
(21). At leading order we may hence assume:
\begin{equation}
\delta(\Omega)=\frac{1}{2} \Omega \left( \frac{1}{v_{g1}}-\frac{1}{v_{g2}} \right)
\;\;, \; \;  \beta(\Omega)=\frac{1}{2} \Omega \left( \frac{1}{v_{g1}}+\frac{1}{v_{g2}}
\right)
\end{equation}
If we introduce the spectral power gain curve $G(\Omega)$ and group delay
$\tau_g(\Omega)$ of the amplifier, defined as $G(\Omega) \equiv |g(\Omega)|^2$ and
$\tau_g(\Omega) \equiv
\partial \phi_g / \partial \Omega$, where $\phi_g(\Omega)$ is the
phase of $g(\Omega)$, from Eqs.(17), (19) and (22), and recalling
that $a(\delta)={\cal M}_{11}(\delta) \exp(-i\delta L)$, one
finally obtains the following relationship between the power
spectral gain $G$ and group delay $\tau_g$ of the amplifier:
\begin{equation}
\tau_g=\frac{L}{v_{g1}}+\frac{1}{2} \left(
\frac{1}{v_{g1}}-\frac{1}{v_{g2}}\right)\frac{1}{ \pi } \int_{-\infty}^{\infty} d
\delta' \frac{\partial {\rm ln}\sqrt {G(\delta')}}{\partial \delta'}
\frac{1}{\delta'-\delta}
\end{equation}
that is:
\begin{equation}
\tau_g=\frac{L}{v_{g1}} \pm \frac{1}{ \pi } \int_{-\infty}^{\infty} d \Omega'
\frac{\partial {\rm ln}\sqrt {G(\Omega')}}{\partial \Omega'} \frac{1}{\Omega'-\Omega}
\end{equation}
where the upper [lower] sign occurs if $v_{g1}< v_{g2}$ [$v_{g1}> v_{g2}$]. Equation
(24) shows that the power spectral gain curve $G(\Omega)$ of the amplifier uniquely
determines its phase response, i.e. the group delay $\tau_g$. The constraint imposed by
Eq.(24) has important impacts on the response of the amplifier, reducing the
flexibility in designing a given spectral gain curve as compared to the related DFG
problem \cite{Imeshev01}. For instance, the possibility of "squaring" the spectral
power gain curve of the amplifier comes at the price of an increased phase distortion
\cite{note1}. Such limitations come out just because the spectral gain curve satisfies
the minimum-phase condition, which is not the case of the DFG transfer function
\cite{note0}. Another interesting consequence of Eq.(24) is the existence of spectral
regions where the group delay becomes superluminal or even negative, which will be
studied in detail in the next section. Here we end our general analysis by briefly
addressing the issue of the synthesis of the spectral gain curve $g(\Omega)$ of a QPM
amplifier (the so-called "inverse problem"), that is the determination of the QPM
profile $d_{eff}(z)$ that realizes a target spectral response $g(\Omega)$. Such a
problem has been extensively studied on several occasions in the context of optical
filters and gratings theory (see, for instance, \cite{Feced99} and references therein),
and therefore it will be briefly quoted here. First of all, let us recall from inverse
scattering theory that in Eq.(14) the scattering potential $q(z)$ is uniquely
determined from the knowledge of $r(\delta)={\cal M}_{12}/{\cal M}_{11}$, which must
satisfy minimal requirements of causality, and that numerical techniques for
determination of $q(z)$ from the knowledge of $r(\delta)$ are well developed
\cite{Feced99,Skaar01}. Let us now assign a target power spectral gain curve
$G(\Omega)$, and ask if there is a QPM grating profile $d_{eff}(z)$ that realizes such
a gain response. First of all, $G(\Omega)$ should satisfy minimal conditions:
$G(\Omega) \geq 1$ for all real frequencies $\Omega$, be analytic in the lower [upper]
part of the complex $\Omega$ plane if $v_g{1}< v_{g2}$ [$v_g{1}> v_{g2}$], $G(\Omega)
\rightarrow 1 $ for $\Omega \rightarrow \infty$, and the integral $\int_{0}^{\infty} d
\Omega{\rm ln}G(\Omega)/(1+ \Omega^2)$ should not diverge. In this case, the scattering
coefficient $a(\delta)$, and hence the matrix coefficient ${\cal M}_{11}$, can be
univocally  determined owing to the minimal phase shift condition through Eq.(19). Once
${\cal M}_{11}(\delta)$ has been calculated from $G(\Omega)$, the modulus of
$r(\delta)$ is determined by $|r(\delta)|^2=(1-1/|{\cal M}_{11}|^2)$, however some
degrees of freedom are left in the phase of $r(\delta)$, yet ensuring the requirements
of causality (see
 Sec.4 of Ref.\cite{Toll56}). By changing the phase of $r(\delta)$, one basically
 changes the spectral {\it phase} response of the generated idler wave, which is related to the phase
 of ${\cal M}_{12}$ (see \cite{note0}), without affecting the spectral
 gain amplitude and phase at the signal wavelength. For a physically
realizable power spectral gain curve $G(\Omega)$, there is not, hence, a unique QPM
profile that realizes such a gain response, the different QPM profiles corresponding to
different spectral phase response of the generated idler wave. The non-uniqueness of
the inverse problem is analogous to that encountered, e.g., in the design of fiber
Bragg gratings for use in transmission \cite{Skaar01bis,note1bis}.

\section{\label{superluminal}Anomalous propagation of spectrally narrow optical pulses:
superluminal and negative transit times}
A physically important consequence of the
dispersive properties of a QPM parametric amplifier, stated by Eq.(24), is the
existence of abnormal group delays, that is the existence of spectral regions where a
spectrally-narrow signal pulse can travel through the amplifier with a superluminal
($\tau_g < L/c_0$) or even negative ($\tau_g<0$) transit time \cite{note2}. Let us
consider the propagation of a spectrally-narrow signal pulse through the QPM amplifier.
We assume that the spectrum $\tilde f(\Omega)$ of the incident pulse is centered at a
frequency $\Omega_0$, so that we can write $A_1(0,t)=\exp(-i \Omega_0 t) h_0(t)$ and
$A_1(L,t)=\exp(-i \Omega_0 t) h_L(t)$ at the input and output planes of the amplifier.
If we now assume that the spectral extent of the envelope $h_0(t)$ is narrow enough
such that the gain curve $g(\Omega)$ of the amplifier varies slowly over the pulse
bandwidth, an asymptotic expansion for $h_L(t)$ may be obtained, as detailed in
Appendix B. At leading order one obtains:
\begin{equation}
h_L(t) \simeq g_0 h_0(t-\Delta \tau)
\end{equation}
where $g_0=g(\Omega_0)$ is the gain of the amplifier at the pulse central frequency
$\Omega_0$ and $\Delta \tau$ is the complex group delay time, which is given by (see
Appendix B):
\begin{equation}
\Delta \tau \equiv -i \left( \frac{\partial {\rm ln} g}{\partial \Omega }
\right)_{\Omega_0}=\tau_g(\Omega_0)-i \left( \frac{\partial {\rm ln}\sqrt G}{\partial
\Omega}\right)_{\Omega_0}.
\end{equation}
Notice that, since the group delay $\Delta \tau$ is in general complex-valued, even at
leading order in the expansion [see Eq.(B7) in Appendix B] the amplifier produces a
pulse distortion \cite{note3}. However, if $\Omega_0$ is a stationary point of the
spectral gain curve, than the complex group delay $\Delta \tau$ becomes real-valued and
equal to the usual group delay $\tau_g=\partial \phi_g / \partial \Omega$. In this
case,  at leading order the amplifier is distortionless and the pulse transit time is
related to the spectral gain curve $G(\Omega)$ by means of the Hilbert transform given
in Eq.(24). Let us now suppose that $v_{g1}>v_{g_2}$ and that $\Omega_0$ is a maximum
of the spectral power gain curve; then for $\Omega \simeq \Omega_0$, the contribution
to the integral on the right hand side in Eq.(24) near the singularity $\Omega'=\Omega$
is negative, and the integral is thus expected to be negative. This implies that
$\tau_g>L/v_{g_1}$, i.e. the effective group velocity of the pulse is lower than
$v_{g1}$. Conversely, if $\Omega_0$ is a minimum of the spectral power gain curve, the
integral in Eq.(23) is expected to be positive, and the transit time $\tau_g$ lower
than $L/v_{g_1}$. This means that superluminal or even negative transit times may be
expected whenever the pulse spectrum is centered around a minimum of the spectral gain
curve \cite{note4}. The existence of anomalous transit times for a pulse tuned midway
two amplification peaks is indeed a rather general result, and previous demonstrations
of negative transit times have been reported for optical pulses as well as for
electronic signals using a gain-doublet amplifier
\cite{Steinberg94,Wang00,Dogariu01,Mitchell97}.\\
 To get a
quantitative analysis of such effects, let us refer to a periodically-poled Lithium
Niobate (PPLN) crystal pumped at the wavelength $\lambda_3=532$ nm with a signal field
at $\lambda_1=1.55 \; \mu$m. The idler wave corresponds to the wavelength
$\lambda_2=810$ nm; we further assume extraordinary wave propagation, so that $d=d_{33}
\simeq 27$ pm/V. The temperature-dependent dispersion relation $k=k(\omega)=n(\omega)
\omega / c_0$ for extraordinary waves in Lithium Niobate is determined using Sellmeir
equations from Ref.\cite{Edwards84}. At 25$^\circ$C, one can estimate $v_{g1} \simeq
0.4815 c_0$, $v_{g2} \simeq 0.44220 c_0$, and a first-order QPM period $\Lambda=7.39 \;
\mu$m, which is accessible with current poling technology. The main task in the design
of the QPM grating is to realize a gain curve $G(\Omega)$ with a local minimum, e.g. at
$\Omega_0=0$.  There are several possibilities to achieve a spectral dip in the gain
curve of a
 QPM amplifier, such as the inclusion of a defect in an otherwise periodic QPM grating or the
 cascading of two periodic QPM gratings. We consider here the latter case, however a
 similar analysis could be done for the former configuration. The QPM grating, shown in
 Fig.2, consists of a sequence of two +/- square-wave uniform gratings, each of length $a$
 and period $\Lambda= 2 \pi / \Delta k$ (first-order QPM), separated by a distance
 $l$. For such a structure one can easily calculate $d_{eff}$, and
 hence $q(z)$, obtaining $q(z)=q_0$ for $0<z<a$, $q(z)=0$ for $a<z<a+l$, and
$q(z)=q_0 \exp(i \Phi) $ for $a+l<z<2a+l$, where $q_0 \equiv (2 / \pi) d_{33} [8 \pi^2
I_3 / (\epsilon_0 c_0 n_1 n_2 n_3 \lambda_1 \lambda_2) ]^{1/2}$ and $\Phi$ is a phase
shift that depends on the relative phases of the two square waves in the two grating
sections. For instance, the reversal of sign of $d_{eff}$, corresponding to $\Phi=\pi$,
occurs if the two square waveforms in the two grating sections are shifted each other
by half a period [see Fig.2(b)]. Notice that, in the limiting case $l \rightarrow 0$,
the structure reduces to a periodic QPM grating of length $2a$ with a defect at the
center of the structure. In order to calculate the spectral gain curve $g(\Omega)$ of
the amplifier, we need to evaluate the matrix element ${\cal M}_{11}$ according to
Eq.(17). The transfer matrix of the structure shown in Fig.2 can be readily calculated
as the product of the three transfer matrices corresponding to the propagation in the
first uniform grating section, the middle grating-free section, and the second uniform
grating section in the reverse order, i.e.:
\begin{eqnarray}
{\cal M}  & = &     \left(
\begin{array}{cc}
\cosh \left( \theta a \right)+i \frac{\delta}{\theta} \sinh \left( \theta a \right) &
i \exp(i \Phi) \frac{q_0}{\theta} \sinh \left( \theta a \right) \\
-i \exp(-i\Phi) \frac{q_0}{\theta} \sinh \left( \theta a \right)&
\cosh \left( \theta a \right)-i \frac{\delta}{\theta} \sinh \left(
\theta a \right)
\end{array}
\right)
 \times
 \left(
\begin{array}{cc}
\exp(i \delta l)  & 0  \\
0 & \exp(-i \delta l)
\end{array}
\right)
 \times \nonumber \\
 & & \left(
\begin{array}{cc}
\cosh \left( \theta a \right)+i \frac{\delta}{\theta} \sinh \left( \theta a \right) & i
\frac{q_0}{\theta}
\sinh \left( \theta a \right) \\
-i \frac{q_0}{\theta} \sinh \left( \theta a \right)& \cosh \left( \theta a \right)-i
\frac{\delta}{\theta} \sinh \left( \theta a \right)
\end{array}
\right)
\end{eqnarray}
where $\theta \equiv \sqrt{|q_0|^2-\delta^2}$. From Eqs.(17) and (27) one readily
obtains:
\begin{equation}
g(\Omega)= \exp(i \delta l+i \beta L  ) \left [  \cosh^2(\theta
a)-\frac{\delta^2}{\theta^2} \sinh^2(\theta a) +i
\frac{\delta}{\theta} \sinh (2 \theta a) + \exp(i\Phi)
\frac{|q_0|^2}{\theta^2} \sinh^2 (\theta a) \exp(-2i \delta l)
\right]
\end{equation}
where $\beta=\beta(\Omega)$ and $\delta=\delta(\Omega)$ are given,
in the general case, by Eqs.(13) and (15).
 Around the resonance $\Omega=0$, the approximate equations (22) may be
used, which require the knowledge of the group velocities for signal and idler fields
solely. A typical behavior of power spectral gain $G(\Omega)$ and corresponding group
delay $\tau_g(\Omega)$ for a few different values of $\Phi$ are shown in Figs.3,4 and
5. The parameter values of the QPM grating structure are $a=3$ mm and $l=2$ mm. Notice
that, in correspondence of local minima in the power spectral gain curve, the group
delay curve shows local minima, corresponding to superluminal and even negative group
delays (see, for instance, Fig.3).  An inspection of Eq.(28) reveals that, for
$\Phi=\pi$, the curve $G(\Omega)=|g(\Omega)|^2$ shows a dip at $\Omega=0$, with
$G(0)=1$ and with a corresponding minimum in the group delay curve which can be
calculated analytically in a closed form and reads:
\begin{equation}
\tau_g(\Omega=0)= \frac{L}{2} \left( \frac{1}{v_{g1}} -\frac{1}{v_{g2}} \right)
-\frac{l}{2}\left( \frac{1}{v_{g2}}-\frac{1}{v_{g1}} \right) \times \left[
\cosh^2(|q_0|a)+\sinh^2(|q_0|a)+ \frac{\sinh(2 |q_0|a)}{|q_0|l}\right]
\end{equation}
 For $\Phi \neq \pi$, the curves $G(\Omega)$ and $\tau_g(\Omega)$
may become strongly asymmetric (as in Fig.4), the minimum in the gain curve is shifted
away from $\Omega=0$ or multiple local minima may occurs; however the general rule of
faster than $v_{g1}$ group velocity near local minima of the power spectral gain curve
holds. Figure 6 shows the behavior of the group delay at $\Omega=0$, for the case of
perfect phase reversal of the two grating sections (i.e., for $\Phi=\pi)$, versus
intensity of the pump wave. The figure clearly indicates the possibility of controlling
the transit time of a spectrally-narrow signal pulse by the pump intensity. Notice that
the group delay becomes superluminal, i.e. smaller than $L/c_0$, at $I_3 \simeq 105$
MW/cm$^2$ and negative at $I_3 \simeq 135$ MW/cm$^2$. To reach the latter regime, the
pump intensity corresponds to an off-resonance power gain peak as large as 40 dB [see
Fig.3(a)]. However, since the pulse spectrum is centered at $\Omega=0$ where $G(\Omega)
\simeq 1$, {\it the main effect of the amplifier on pulse propagation near the gain dip
is to advance the pulse in time, not to amplify it}. A similar effect was predicted and
observed for pulse propagation in an atomic gain-doublet amplifier
\cite{Wang00,Dogariu01}. What physically happens is that in the fist section of the QPM
grating the signal pulse is amplified, with the generation of the idler wave ($\omega_3
\rightarrow \omega_1+\omega_2)$; however owing to the phase reversal in the second QPM
grating, a back conversion process ($\omega_1+\omega_2 \rightarrow \omega_3$) occurs in
the second grating section. The result of such a cascading process is that the signal
pulse is basically not amplified at the output of the crystal, however the phase delays
suffered by its spectral components, as ruled by the causality condition [Eq.(24)],
produce a temporal advancement with no appreciable pulse distortion. We have checked
the prediction of pulse propagation based on the group delay analysis by direct
numerical simulation of pulse propagation in a two-section QPM PPLN crystal with
perfect phase reversal ($\Phi=\pi$) starting from Eqs.(8) in the limit of an undepleted
pump. As a probing pulse, we assumed a transform-limited Gaussian pulse, tuned at
$\Omega=0$, with a pulse duration (FWHM) of 250 ps to ensure the spectrally-narrow
pulse limit. Figure 7 shows the traces of the incident signal pulse at the entrance
plane of the crystal $z=0$ (dotted line) and of transmitted pulses at the exit plane
$z=L$ (solid lines) for increasing values of the pump intensity $I_3$; the pulse
intensity of the output waveforms are normalized to the peak intensity of incident
pulse. It is remarkable that, at pump intensities corresponding to, e.g., curve 4, the
transmitted pulse leaves the amplifier {\it before} the peak of Gaussian incident pulse
has entered into the crystal. To make a quantitative estimate of pump power levels
required to observe such effects, let us assume a Gaussian pump with a beam waist of
$\simeq 200 \; \mu$m; then curve 3 of Fig.7 corresponds to a pump peak power of $\simeq
170$ kW. Using a pulsed pump of duration (FWHM) of $\sim$ 5 ns, i.e. about twenty times
longer than the probing pulses, a pump pulse energy of $\simeq 0.90$ mJ is required,
which can be obtained using a frequency-doubled Q-switched Nd-based laser system as a
pump source.

\section{Conclusions}
In this paper we have presented a general analysis of the dispersive properties of QPM
optical parametric amplifiers. In the limit of no pump depletion, the parametric
interaction of idler and signal fields, that accounts for material dispersion at any
order, may be described by coupled-mode equations which have a canonical form widely
encountered in problems of inverse scattering and optical gratings and filters design.
One of the main consequences of the analysis is that the spectral power gain curve of
the amplifier defines univocally its dispersive curve through a Hilbert-like transform.
This circumstance may be of major relevance in the design and synthesis of QPM
amplifiers for applications in pulse shaping and control, and imposes unavoidable
physical limits to the realization of dispersionless amplifiers with a flat spectral
gain \cite{note1}. Our analysis also demonstrates that engineered QPM grating profiles
can simulate dispersive properties of resonant pulse propagation in inverted atomic
media. In particular, in this work we have proposed and studied the possibility of
observing and controlling superluminal and negative group velocities of picosecond
optical pulses in a PPLN amplifier, which shows features similar to those found in
atomic amplifiers with a gain doublet \cite{Steinberg94,Wang00,Dogariu01}.


\appendix

\appendix
\section{Derivation of envelope equations}
In this Appendix we derive the envelope equations [Eqs.(8a)-(8c)] given in the text by
a multiple scale asymptotic analysis of the original equations [Eqs.(3a)-(3c)]. Such an
analysis assumes that the nonlinearity of the medium is weak and that the spectral
extent of the interacting fields is narrow enough such that in Eq.(5) the operator
$k^2(\omega_0+i\partial_t)-k^2(\omega_0)$ can be treated as a perturbation term. With
these assumptions, it is worth rewriting Eqs.(3a)-(3c) in the form:
\begin{equation}
\frac{\partial^2 {\cal E}_1}{\partial z^2}+k_{1}^{2} {\cal E}_1 = \epsilon \left\{
\left[ k_{1}^{2}-k^2(\omega_1+i \partial_t) \right] {\cal E}_1-\chi^{(2)}
\left(\frac{\omega_1}{c_0} \right)^2 {\cal E}_3 {\cal E}_{2}^{*} \right\}
\end{equation}
\begin{equation}
\frac{\partial^2 {\cal E}_2}{\partial z^2}+k_{2}^{2} {\cal E}_2 = \epsilon \left\{
\left[ k_{2}^{2}-k^2(\omega_2+i \partial_t) \right] {\cal E}_2-\chi^{(2)}
\left(\frac{\omega_2}{c_0} \right)^2 {\cal E}_3 {\cal E}_{1}^{*} \right\}
\end{equation}
\begin{equation}
\frac{\partial^2 {\cal E}_3}{\partial z^2}+k_{3}^{2} {\cal E}_3 = \epsilon \left\{
\left[ k_{3}^{2}-k^2(\omega_3+i \partial_t) \right] {\cal E}_3-\chi^{(2)}
\left(\frac{\omega_3}{c_0} \right)^2 {\cal E}_1 {\cal E}_{2} \right\}
\end{equation}
where $\epsilon$ is a small parameter that defines the order of
magnitude of the perturbation terms entering on the right hand
side in Eqs.(A1)-(A3). The problem is to construct an asymptotic
solution of the perturbed equations as $\epsilon \rightarrow 0$
which is valid uniformly with respect to the spatial variable $z$.
Therefore we look for a solution to Eqs.(A1)-(A3) in the form:
\begin{equation}
{\cal E}_l={\cal E}_{l}^{(0)}+\epsilon {\cal E}_{l}^{(1)}+ \epsilon^2 {\cal
E}_{l}^{(2)}+ ...
\end{equation}
($l=1,2,3$), and we require that the asymptotic expansion be uniformly valid. This
condition can be satisfied by introducing multiple scales for $z$, i.e. by assuming
that ${\cal E}_l$ depends on $Z_0$, $Z_1$, $Z_2$, ..., where $Z_0=z$, $Z_1= \epsilon
z$, $Z_2= \epsilon^2 z$, ... The introduction of multiple spatial scales is fundamental
to remove secular growing terms that arise in the perturbation expansion. Introducing
expansion (A4) into Eqs.(A1)-(A3), using the derivative rule $\partial^{2}_{z}=
\partial^{2}_{Z_0}+ \epsilon(\partial_{Z_0} \partial_{Z_1}+\partial_{Z_1}
\partial_{Z_0})+...$ and collecting the terms of the same order in the equations so
obtained, a hierarchy of equations for successive corrections to ${\cal E}_l$ is
obtained. At leading order, $O(\epsilon^0)$, one obtains:
\begin{equation}
\frac{\partial^2 {\cal E}_{l}^{(0)}}{\partial Z_{0}^{2}}+k_{l}^{2} {\cal E}_{l}^{(0)}=0
.
\end{equation}
($l=1,2,3$). If we consider forward propagating waves, i.e. a copropagating
interaction, the solutions to Eq.(A5) are given by:
\begin{equation}
{\cal E}_{l}^{(0)}(Z_0,Z_1,...;t)=A_l(Z_1,...;t) \exp(ik_l Z_0) ,
\end{equation}
where the amplitudes $A_l$ depend on time and on slow spatial
variables. At $O(\epsilon)$ one obtains:
\begin{equation}
\frac{\partial^2 {\cal E}_{l}^{(1)}}{\partial Z_{0}^{2}}+k_{l}^{2}
{\cal E}_{l}^{(1)}=G^{(1)}_{l}
\end{equation}
where the driving terms $G_{l}^{(1)}$ in Eqs.(A7) are given by:
\begin{equation}
G^{(1)}_{1}=-2 \partial_{Z_0} \partial_{Z_1} {\cal E}^{(0)}_{1}+ \left[
k_{1}^{2}-k^2(\omega_1+i \partial_t) \right] {\cal E}^{(0)}_{1} -\chi^{(2)} \left(
\frac{\omega_1}{c_0} \right)^2 {\cal E}_{3}^{(0)} {\cal E}_{2}^{(0)*}
\end{equation}
\begin{equation}
G^{(1)}_{2}=-2 \partial_{Z_0} \partial_{Z_1} {\cal E}^{(0)}_{2}+ \left[
k_{2}^{2}-k^2(\omega_2+i \partial_t) \right] {\cal E}^{(0)}_{2} -\chi^{(2)} \left(
\frac{\omega_2}{c_0} \right)^2 {\cal E}_{3}^{(0)} {\cal E}_{1}^{(0)*}
\end{equation}
\begin{equation}
G^{(1)}_{3}=-2 \partial_{Z_0} \partial_{Z_1} {\cal E}^{(0)}_{3}+ \left[
k_{3}^{2}-k^2(\omega_3+i \partial_t) \right] {\cal E}^{(0)}_{3} -\chi^{(2)} \left(
\frac{\omega_3}{c_0} \right)^2 {\cal E}_{1}^{(0)} {\cal E}_{2}^{(0)}.
\end{equation}
To further proceed, let us assume that $\chi^{(2)}(z)$ is a quasi-periodic function of
$z$ with a period $\Lambda$ satisfying the QPM condition given by Eq.(7) in the text,
and that $\Lambda$ is much smaller than the crystal length and of the same order of
magnitude as the wavelengths of interacting fields. It is then worth separating the
fast and slow dependence of $\chi^{(2)}$ on $z$ by setting:
\begin{equation}
\chi^{(2)}(z)=\sum_{n=-\infty}^{\infty} \chi^{(2)}_{n}(Z_1) \exp(-2 \pi i n Z_0 /
\Lambda)
\end{equation}
where the coefficients $\chi^{(2)}_{n}$  of the Fourier series are allowed to vary on
the slow spatial scale $Z_1$ and satisfy the condition
$\chi^{(2)}_{-n}=\chi^{(2)*}_{n}$. The solutions to Eqs. (A7) are bounded with respect
to $Z_0$ provided that the driving term $G_{l}^{(1)}$ does not contain terms
oscillating like $\exp(ik_lZ_0)$. The solvability conditions allow one to derive the
evolution equations of the envelopes $A_l$ on the slow spatial scale $Z_1$.
Substitution of Eqs.(A6) and (A11) into Eqs.(A8)-(A10), after collecting the the terms
oscillating like $\exp(ik_lZ_0)$, one obtains the following solvability conditions:
\begin{equation}
-2ik_1 \frac{\partial A_1}{\partial Z_1}+
\left[k_{1}^{2}-k^2(\omega_1+ i \partial_t) \right] A_1 -
\chi^{(2)}_{M} \left( \frac{\omega_1}{c_0} \right)^2 A_{2}^{*} A_3
=0
\end{equation}
\begin{equation}
-2ik_2 \frac{\partial A_2}{\partial Z_1}+
\left[k_{2}^{2}-k^2(\omega_2+ i \partial_t) \right] A_2 -
\chi^{(2)}_{M} \left( \frac{\omega_2}{c_0} \right)^2 A_{1}^{*} A_3
=0
\end{equation}
\begin{equation}
-2ik_3 \frac{\partial A_3}{\partial Z_1}+ \left[k_{3}^{2}-k^2(\omega_3+ i \partial_t)
\right] A_3 - \chi^{(2)*}_{M} \left( \frac{\omega_3}{c_0} \right)^2 A_1 A_2 =0
\end{equation}
If we stop the asymptotic expansion at order $\sim \epsilon$,
re-introducing the original spatial variable $z$ instead of $Z_1$
in Eqs.(A12)-(A14) and setting $\epsilon=1$, one finally obtains
Eqs.(8) given in the text, where $d_{eff}$ is defined through
Eq.(9).

\section{Propagation of spectrally-narrow optical pulses}

Let $A_1(0,t)=\exp(-i \Omega_0 t) h_0(t)$ be the signal field envelope incident upon
the amplifier with a frequency offset $\Omega_0$ from the carrier $\omega_1$. From
Eq.(11) the pulse waveform at the exit of the amplifier can be written as
$A_1(L,t)=h_L(t) \exp(-i \Omega_0 t)$, where:
\begin{equation}
h_L(t)=g(\Omega_0+i \partial_t) h_0(t)
\end{equation}
and the operator on the right hand side in Eq.(B1) is defined, as usual, by its power
series expansion \cite{Newell92}.  We now assume that the spectral extent of the
envelope $h_0(t)$ is narrow enough such that the gain curve $g(\Omega)$ of the
amplifier varies slowly over the pulse bandwidth. One can then expand $g(\Omega)$ in
power series around $\Omega=\Omega_0$ as follows:
\begin{eqnarray}
g(\Omega)=\exp[{\rm ln} g(\Omega)]= g_0 \exp \left[ \left( \frac{\partial {\rm ln} g}
{\partial \Omega} \right)_{\Omega_0} (\Omega-\Omega_0) + \right. \nonumber \\
\left. \frac{1}{2}\left( \frac{\partial^2 {\rm ln} g} {\partial \Omega^2}
\right)_{\Omega_0} (\Omega-\Omega_0)^2+ ... \right]=g_0{\cal B} \exp \left[ \left(
\frac{\partial {\rm ln} g} {\partial \Omega} \right)_{\Omega_0} (\Omega-\Omega_0)
\right]
\end{eqnarray}
where we have set:
\begin{equation}
{\cal B}(\Omega-\Omega_0)= \exp \left[ \frac{1}{2}\left( \frac{\partial^2 {\rm ln} g}
{\partial \Omega^2} \right)_{\Omega_0} (\Omega-\Omega_0)^2+ ... \right].
\end{equation}
Substitution of Eq.(B2) into Eq.(B1), after observing that the
operator $ \exp(- \Delta \tau \partial_t)$ is equivalent to the
time translation $t \rightarrow t-\Delta \tau$, one obtains:
\begin{equation}
h_L(t)=g_0{\cal B}( i \partial_t) h_0(t-\Delta \tau)
\end{equation}
where we have introduced the "complex" group delay:
\begin{equation}
\Delta \tau \equiv -i \left( \frac{\partial {\rm ln} g}{\partial \Omega }
\right)_{\Omega_0}=\tau_g(\Omega_0)-i \left( \frac{\partial {\rm ln}\sqrt G}{\partial
\Omega}\right)_{\Omega_0}.
\end{equation}
From Eq.(B3), the following asymptotic expansion holds for the operator ${\cal B}(i
\partial_t)$:
\begin{equation}
{\cal B}(i \partial_t)=1-\frac{1}{2} \left( \frac{\partial^2 {\rm ln}g}{\partial
\Omega^2}\right)_{\Omega_0} \frac{\partial^2}{\partial t^2} +...
\end{equation}
where the dots involve higher-order time derivatives; substitution of Eq.(B6) into
Eq.(B4) finally yields:
\begin{equation}
h_L(t)=g_0 h_0(t-\Delta \tau)-\frac{g_0}{2} \left( \frac{\partial^2 {\rm ln}g}{\partial
\Omega^2}\right)_{\Omega_0} \frac{\partial^2 h_0}{\partial t^2}(t-\Delta \tau)+...
\end{equation}
At the leading order in the expansion, Eq.(B7) reduces to Eq.(25) given in the text.

\newpage

{\bf Figure Captions}\\
\\
{\bf Fig.1} Schematic of parametric amplification of a weak signal wave at frequency
$\omega_1$ in a QPM nonlinear crystal pumped by a strong CW pump at frequency
$\omega_3$ (copropagating interaction).\\
\\
{\bf Fig.2} PPLN crystal geometry. In (a): the crystal comprises two uniform grating
sections, each of length $a$, separated by a distance $l$ with no grating structure. In
each grating section, a +/- square periodic wave pattern of domains with period
$\Lambda$ is assumed. This yields $d_{eff}=(2/ \pi) d_{33}$ in the first grating and
$d_{eff}=\exp(i\Phi) (2 / \pi) d_{33}$ in the second grating, where the phase $\Phi$ is
determined by the relative phase shift of square waves in the two grating sections. In
(b) it is shown the QPM square wave profile $\chi^{(2)}(z)$ that corresponds to $\Phi=\pi$.\\
\\
{\bf Fig.3} Behavior of spectral power gain (a) and group delay (b) for $\Phi=\pi$. The
pump intensity is $I_3=135$ MW/cm$^2$. The other parameter values are given in the
text. Solid and dashed curves refer to the results obtained by use of Sellmeier
equations, i.e. taking into account dispersion at any order, and by use of approximate
equations (22) given in the text, respectively. The inset in (b) shows an enlargement
of the group delay near resonance.\\
\\
{\bf Fig.4} Same as Fig.3, but for $\Phi=\pi /2$.\\
\\
{\bf Fig.5} Same as Fig.3, but for $\Phi=0$.\\
\\
{\bf Fig.6} Behavior of group delay $\tau_g$ versus pump intensity $I_3$ at the center
of the amplifier gain dip. Parameter values are the same as in Fig.3.\\
\\
{\bf Fig.7} Traces of the intensity of transmitted signal pulse (solid curves) at the
exit plane $z=L$ of the amplifier for a few values of pump intensity. Curve 1: $I_3=0$;
curve 2: $I_3=108$ MW/cm$^2$;  curve 3: $I_3=135$ MW/cm$^2$;  curve 4: $I_3=162$
MW/cm$^2$. The dashed curve is the trace of incident Gaussian pulse at the input plane
$z=0$.

\newpage
\begin{figure}
\vspace*{5cm}
\includegraphics{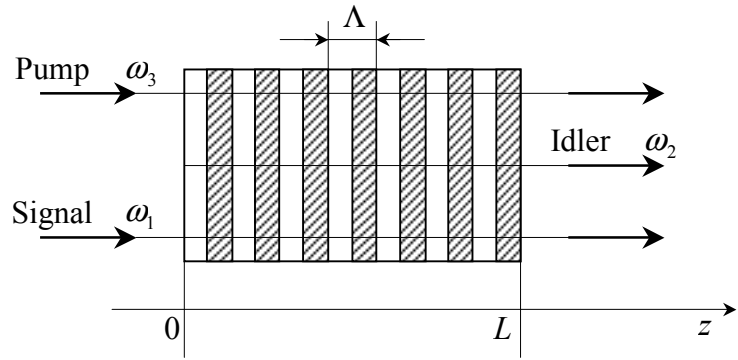}
\vspace*{5cm} \caption{S. Longhi et al., "Dispersive properties of ..."}
\end{figure}

\begin{figure}
\vspace*{5cm}
\includegraphics{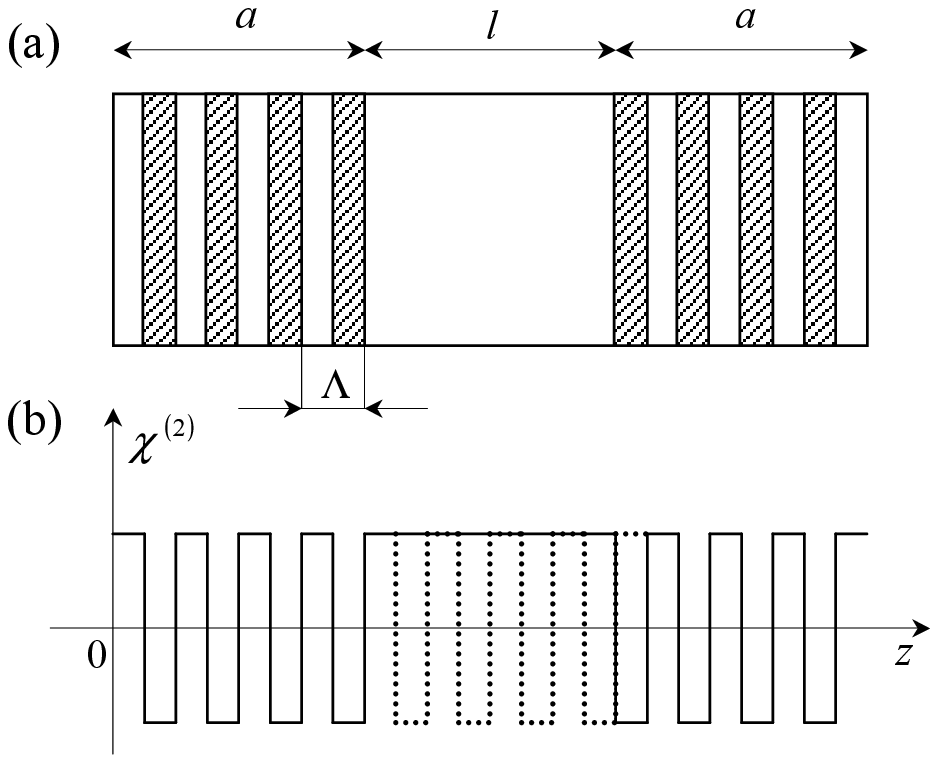}
\vspace*{5cm} \caption{S. Longhi et al., "Dispersive properties of ..."}
\end{figure}

\newpage
\begin{figure}
\vspace{2cm}
\includegraphics{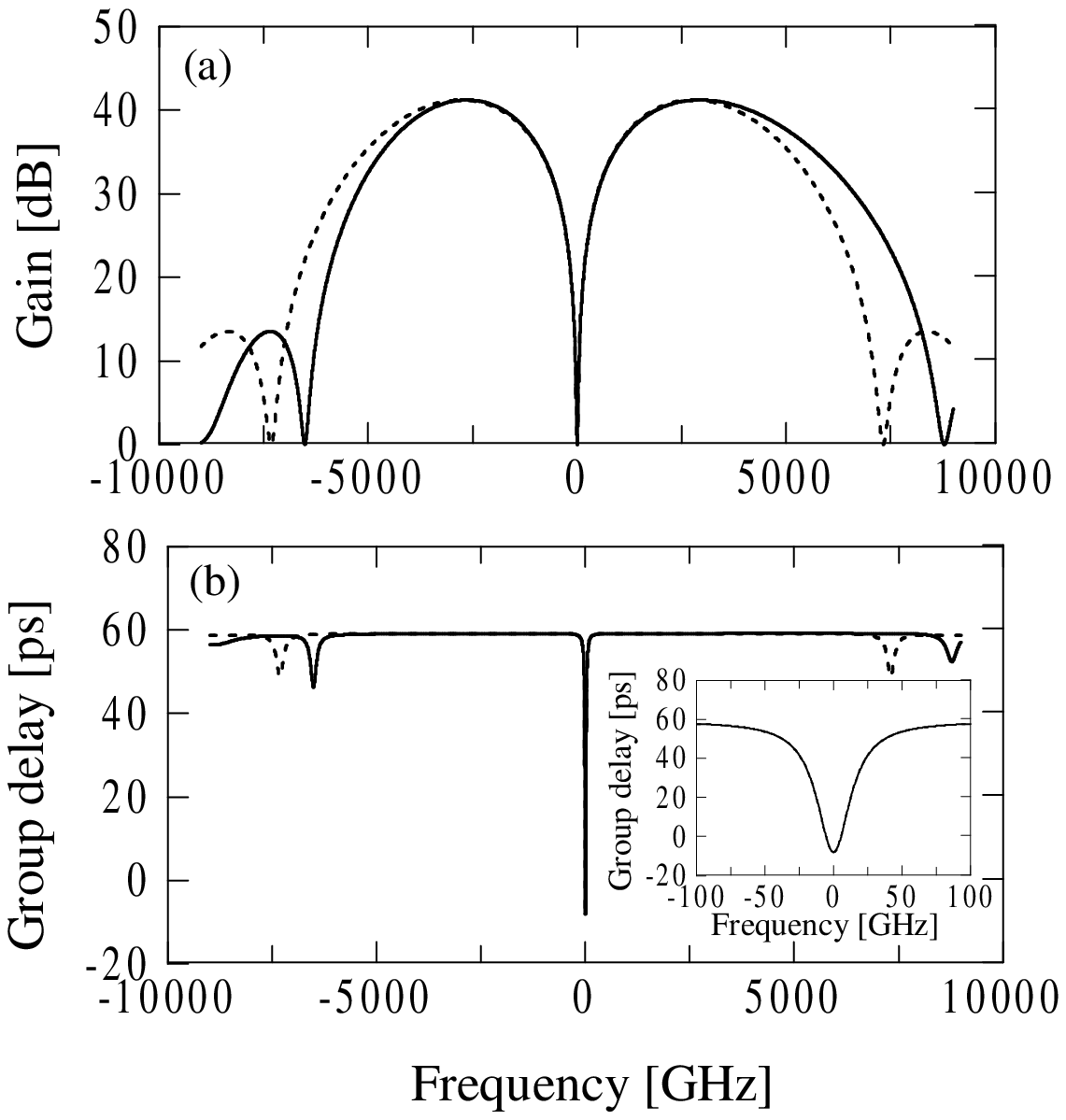}
\vspace{5cm}
\caption{S. Longhi et al., "Dispersive properties of ..."}
\end{figure}

\newpage
\begin{figure}
\vspace{2cm}
\includegraphics{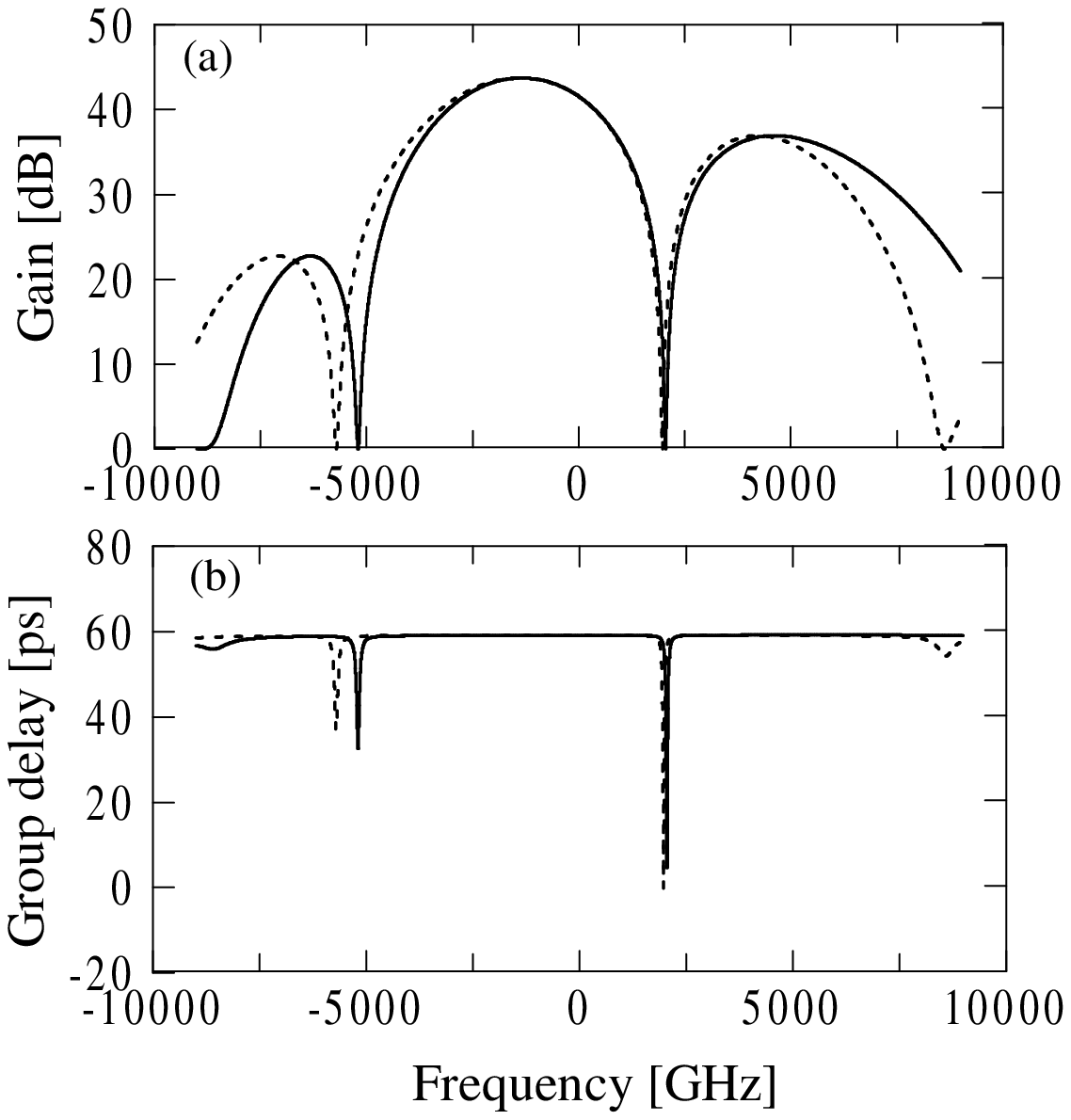}
\vspace{5cm}
\caption{S. Longhi et al., "Dispersive properties of ..."}
\end{figure}

\newpage
\begin{figure}
\vspace{2cm}
\includegraphics{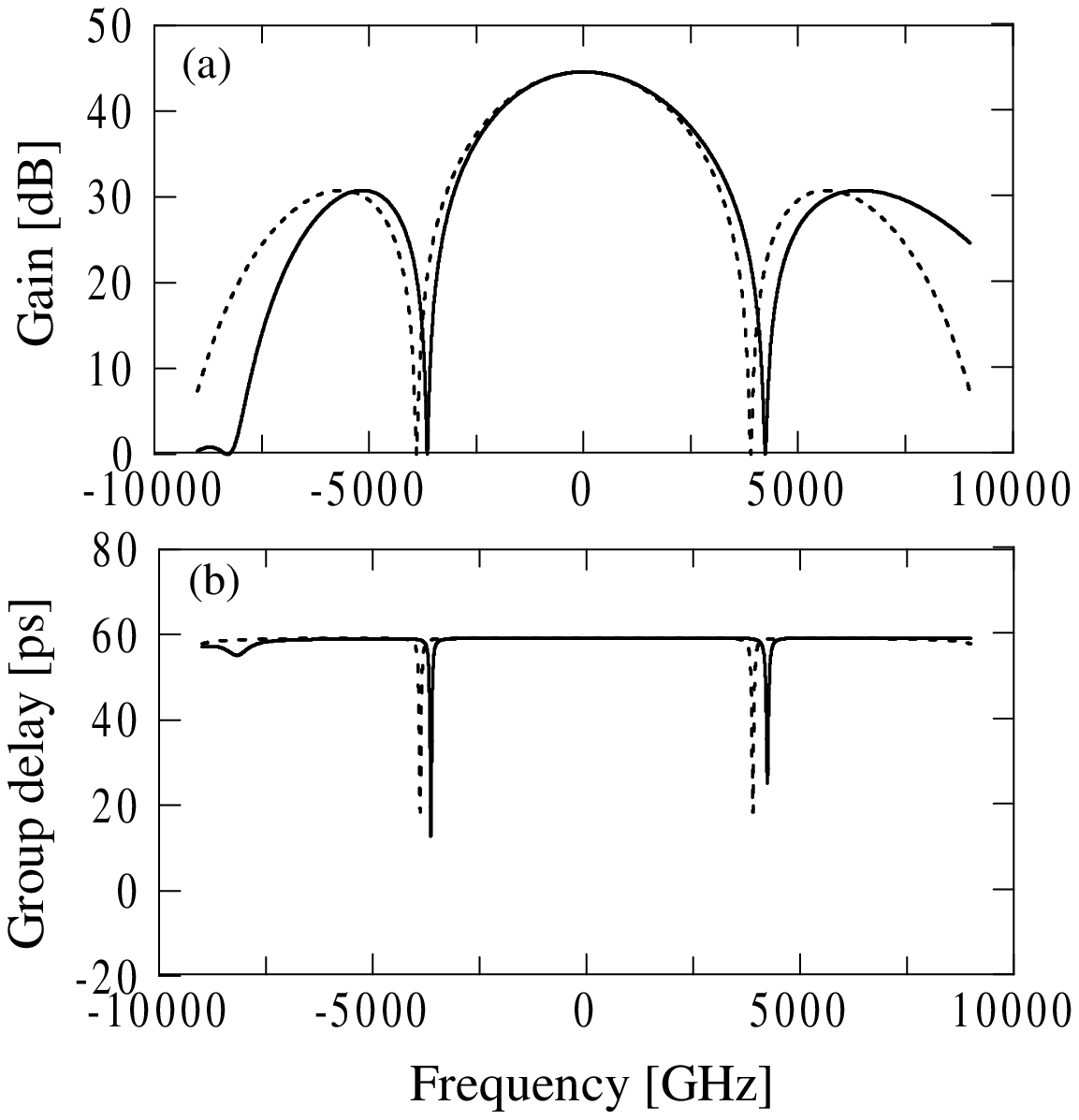}
\vspace{5cm}
\caption{S. Longhi et al., "Dispersive properties of ..."}
\end{figure}

\newpage
\begin{figure}
\vspace{5cm}
\includegraphics{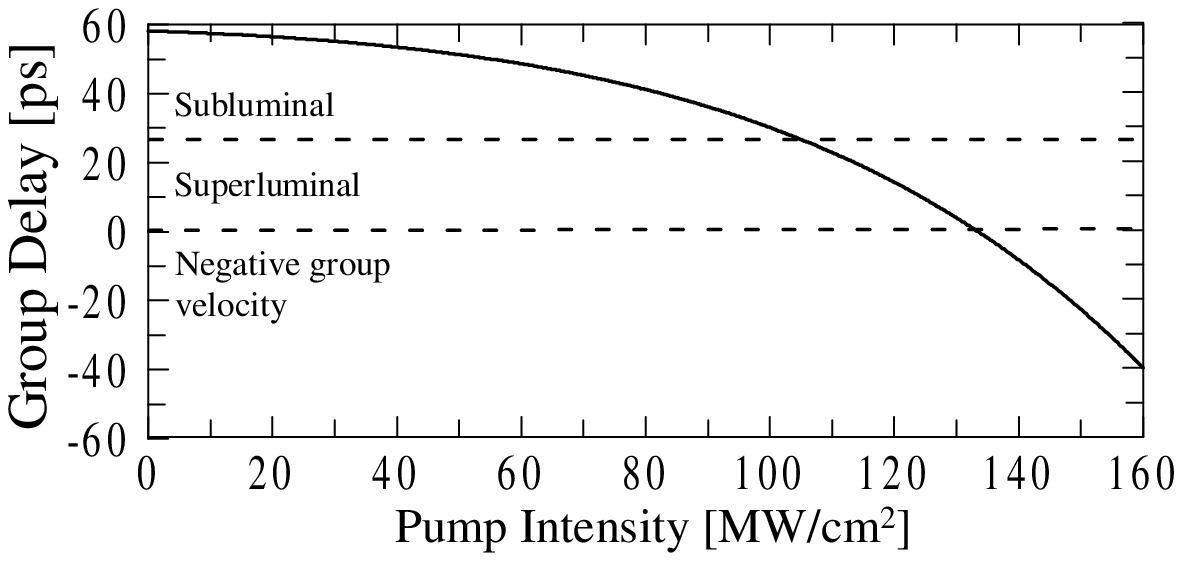}
\vspace{5cm}
\caption{S. Longhi et al., "Dispersive properties of ..."} \vspace{5cm}
\end{figure}

\newpage
\begin{figure}
\vspace{5cm}
\includegraphics{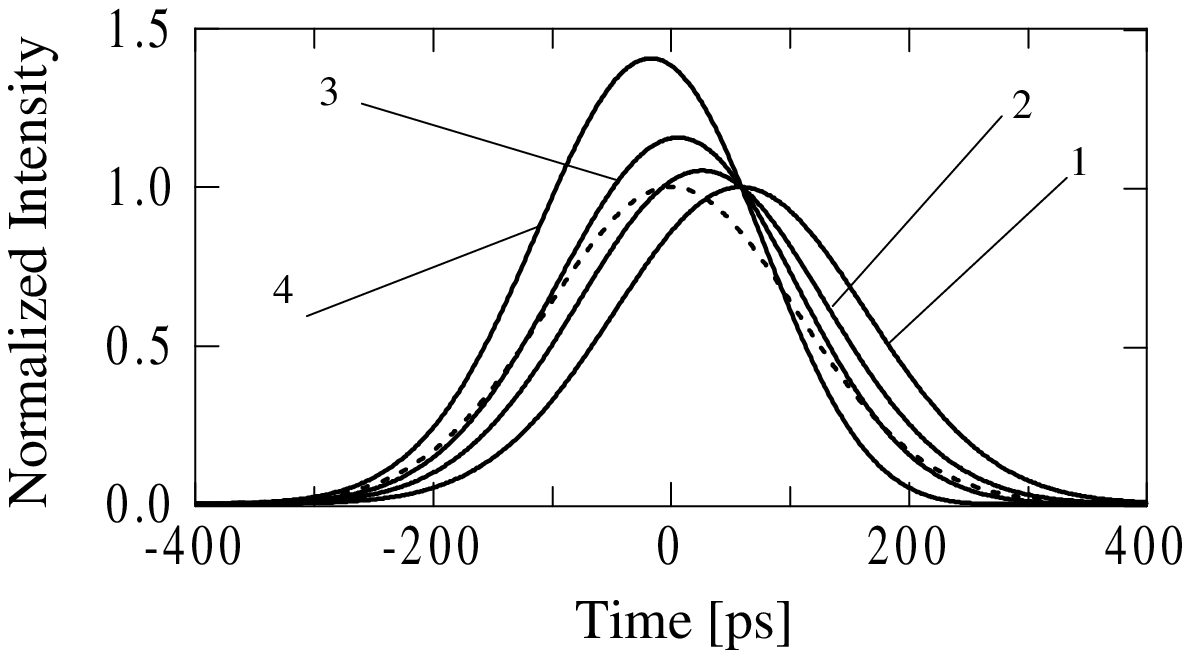}
\vspace{5cm} \caption{S. Longhi et al., "Dispersive properties of ..."}
\end{figure}


\begin{thebibliography}{}
\bibitem{note0}
We note that, using Eq.(12b) and the property ${\cal M}_{21}^{*}={\cal M}_{12}$, one
realizes at once that the generated idler wave at the output plane $z=L$ of the
amplifier is given by: $A_2(L,t)=\int_{-\infty}^{\infty} d \Omega s(\Omega)\tilde
f(\Omega) \exp(i \Omega t)$, where the spectral transfer function $s(\Omega)$ for the
DFG process is given by: $s(\Omega)= (n_1 / n_2) (k_2 / k_1)^{1/2} \exp[-i
\beta(\Omega)L] {\cal M}_{12}(\Omega)$.
\bibitem{note1bis}
A possible profile may be obtained by imposing the minimal phase
shift requirement for $r(\delta)$, that is by choosing the phase of $r(\delta)$ using
again a Hilbert transform. More complex methods, that enable greater flexibility in the
design, have been recently discussed in \cite{Skaar01bis}.
\bibitem{note4}
For the case $v_{g_1}<v_{g2}$, the reverse argument may be applied, i.e. the effective
group velocity becomes larger than $v_{g1}$ when the pulse spectrum is tuned at a local
maximum of the gain curve. In this case, however, superluminal or negative transit
times are unlikely. For instance, for a periodic QPM amplifier of length $L$
[$q(z)=q_0$ constant], the peak of the gain curve occurs at $\Omega_0=0$ and it is
given by $G_0=\cosh^2(q_0L)$; the corresponding group delay can be analytically
calculated and reads
$\tau_g=(L/2)(1/v_{g1}+1/v_{g2})+(L/2)(1/v_{g1}-1/v_{g2})[\tanh(q_0L)/(q_0L)]$. The
effective group velocity, $L/\tau_g$, is larger than $v_{g1}$, however it is always
smaller than $v_{g2}$.
\end{thebibliography}
\end{document}